\documentclass[preprint,showpacs,preprintnumbers,amsmath,amssymb,nofootinbib]{revtex4}

\usepackage{booktabs}
\usepackage{mathrsfs}
\usepackage{epsfig}
\usepackage{graphicx}% Include figure files
\usepackage{dcolumn}% Align table columns on decimal point
\usepackage{bm}% bold math
\usepackage{amsmath}
\usepackage{slashed}
\usepackage{multirow}
\usepackage{subfigure}
\usepackage{epstopdf}
\usepackage{color}

\let\jnfont=\rm
\def\NPB#1,{{\jnfont Nucl.\ Phys.\ B }{\bf #1},}
\def\PLB#1,{{\jnfont Phys.\ Lett.\ B }{\bf #1},}
\def\EPJC#1,{{\jnfont Eur.\ Phys.\ Jour.\ C }{\bf #1},}
\def\PRD#1,{{\jnfont Phys.\ Rev.\ D }{\bf #1},}
\def\PRL#1,{{\jnfont Phys.\ Rev.\ Lett.\ }{\bf #1},}
\def\MPLA#1,{{\jnfont Mod.\ Phys.\ Lett.\ A }{\bf #1},}
\def\JPG#1,{{\jnfont J.\ Phys.\ G}{\bf #1},}
\def\CTP#1,{{\jnfont Commun.\ Theor.\ Phys.\ }{\bf #1},}
\def\ZPC#1,{{\jnfont Z.\ Phys.\ C }{\bf #1},}
\def\JHEP#1,{{\jnfont JHEP \ }{\bf #1},}
\def\Rv{\not{\hbox{\kern-1pt $R$}}}
\def\p{\not{\hbox{\kern-3pt $p$}}}

\newcommand{\bea}{\begin{eqnarray}}
\newcommand{\eea}{\end{eqnarray}}

\newcommand{\bcen}{\begin{center}}
\newcommand{\ecen}{\end{center}}

\newcommand{\ee}{e^+e^-}

\newcommand{\beq}{\begin{eqnarray}}
\newcommand{\eeq}{\end{eqnarray}}

\def\t1{\tilde{t_1}}

\def\be{\begin{equation}}
\def\ee{\end{equation}}
\def\bea{\begin{array}}
\def\eea{\end{array}}
\def\beqa{\begin{eqnarray}}
\def\eeqa{\end{eqnarray}}
\def\beqas{\begin{eqnarray*}}
\def\eeqas{\end{eqnarray*}}

\def\bp{\begin{picture}}
\def\ep{\end{picture}}
\def\bc{\begin{center}}
\def\ec{\end{center}}
\def\bfig{\begin{figure}}
\def\efig{\end{figure}}

\def\bit{\begin{itemize}}
\def\eit{\end{itemize}}

\def\f{\frac}

\def\[{\left[}
\def\]{\right]}
\def\({\left(}
\def\){\right)}

\def\..{\left.}
\def\.{\right.}

\def\ep{\epsilon}

\begin{document}

\preprint{IPMU18-0156}

\title{Vacuum stability in stau-neutralino coannihilation in MSSM}

\author{Guang Hua Duan$^{1,2}$, Chengcheng Han$^{3}$,  Bo Peng$^{1,2}$, Lei Wu$^{4}$, Jin Min Yang$^{1,2,5}$}
%\vspace*{.5cm}}
\affiliation{
$^1$ CAS Key Laboratory of Theoretical Physics, Institute of Theoretical Physics, Chinese Academy of Sciences, Beijing 100190, China \\
$^2$ School of Physical Sciences, University of Chinese Academy of Sciences, Beijing 100049, China \\
$^3$ Kavli IPMU (WPI), UTIAS, University of Tokyo,  Kashiwa, Chiba 277-8583, Japan  \\
$^4$ Department of Physics and Institute of Theoretical Physics, Nanjing Normal University, Nanjing 210023, China \\
$^5$ Department of Physics, Tohoku University, Sendai 980-8578, Japan
}

\begin{abstract}
The stau-neutralino coannihilation provides a feasible way to accommodate the observed cosmological
dark matter (DM) relic density in the minimal supersymmetric standard model (MSSM).
In such a coannihilation mechanism the stau mass usually has an upper bound since its annihilation rate
becomes small with the increase of DM mass. Inspired by this observation, we examine the upper limit
of stau mass in the parameter space with a large mixing of staus. We find that the stau pair
may dominantly annihilate into dibosons and hence the upper bound on the stau mass ($\sim400$ GeV)
obtained from the $f\bar{f}$ final states can be relaxed. Imposing the DM relic density constraint
and requiring a long lifetime of the present vacuum, we find that the lighter stau mass can be as
heavy as about 1.4 TeV for the stau maximum mixing. However, if requiring the present vacuum to
survive during the thermal history of the universe, this mass limit will reduce to about 0.9 TeV.
We also discuss the complementarity of vacuum stability and direct detections in probing this
stau coannihilation scenario.

\end{abstract}
\maketitle

\section{introduction}
It is well known that about 27\% of the global energy budget is dark matter (DM).
So far a lot of experimental efforts have been devoted to DM, such as direct detections, indirect
detections and collider searches, but its nature still remains elusive. The weakly interacting
massive particle (WIMP) is one of the most competitive DM candidates. The lightest neutralino
$\chi^0_1$ in the MSSM can serve as WIMP dark matter naturally if the $R$-parity is conserved.

In general, there exist two generic mechanisms to obtain the correct DM relic
density~\cite{Kolb1990,Griest:1990kh,Jungman:1995df}. One is that the lightest supersymmetric
particle (LSP) can pairly annihilate into the SM particles. If the LSPs are wino or higgsino-like,
their mass should be heavier than 1 TeV because their annihilation rate is large.
If the LSPs are bino-like, they need to annihilate through $Z$ or Higgs funnels or mix with
higgsinos/winos to avoid overclosing
the universe~\cite{Baer:1995nc,ArkaniHamed:2006mb,Abdughani:2017dqs,Pozzo:2018anw}.
The other is that the LSP is bino-like and co-annihilates with some other species
(e.g. a stop~\cite{Boehm:1999bj}, a stau~\cite{Ellis:1998kh}, a wino~\cite{Baer:2005jq},
or gluino~\cite{Profumo:2004wk}) when their masses are nearly degenerate.
In such coannihilation scenarios, the light sparticles are still allowed by the LHC direct searches
because they usually produce soft objects in the final states and are difficult
to be observed at colliders. The delicate searches have been proposed to probe these compressed
scenarios~\cite{Han:2014xoa,Profumo:2017ntc,Duan:2018rls, Han:2015lha, Aboubrahim:2017aen,  Han:2013usa}.
On the other hand, with the increase of the LSP mass, the coannihilation rate becomes small
so that the observed relic density will produce an upper limit on the mass of LSP and also its
co-annihilating partner.

In this work, we focus on the stau-neutralino coannihilation in a simplified MSSM and
attempt to investigate the upper limit of stau mass under the available constraints.
Due to the large tau Yukawa contributions to the renormalization group evolution,
the staus in some high-scale SUSY models tend to be lighter than other sleptons and
may co-annihilate with the neutralino DM. This was first noticed in~\cite{Ellis:1998kh}
and then was calculated in details in~\cite{Ellis:1999mm,Belanger:2001fz,Baer:2002fv,Nihei:2002sc}.
Note that the stau coannihilation is usually dominated by the process
$\tilde{\tau} \tilde{\tau}^* \to f \bar{f}$~\cite{Ellis:1998kh,Han:2016gvr,Athron:2017qdc},
which requires the mass of stau $\lesssim400$ GeV to satisfy the observed DM relic
density\footnote{A recent work on multi-slepton coannhilation without mixing can be
found in \cite{Baker:2018uox}. }. While in the parameter space with a large mixing,
the staus will mainly annihilate into dibosons, $\tilde{\tau} \tilde{\tau}^* \to hh/ZZ/W^+ W^-$,
due to the enhanced couplings between stau and $h/Z/W^\pm$. This may lift the previous upper bound
on the stau mass in the coannihilation. However, such a large mixing of staus may induce
a new charge-breaking vacuum and affect the vacuum stability because of the tunneling effect
~\cite{Casas:1995pd,Baer:1996jn,Kusenko:1996jn,Dasgupta:1998qy,Kitahara:2012pb,Carena:2012mw, Hisano:2010re, Kitahara:2013lfa,Blinov:2013fta,Chowdhury:2013dka,Bobrowski:2014dla,Hollik:2016dcm}. Therefore, it is meaningful
to explore the vacuum stability constraint on the stau sector in the stau coannihilation scenario.

The structure of this paper is organized as follows. In Section~\ref{section2}, we recapitulate the
calculations of DM relic density of coannihilation and the vacuum stability in the MSSM.
In Section~\ref{section3}, we present the numerical results and discussions.
Finally, we draw our conclusions in Section~\ref{section4}.

\section{Stau Coannihilation and Vacuum stability}\label{section2}
The time evolution of a stable particle is described by Boltzmann equation
\begin{equation}\label{Beq}
\frac{dn}{dt} = -3Hn - \langle\sigma v\rangle[n^{2}-(n^{ eq})^{2}]
\end{equation}
where $n$ ($n^{ eq}$) denotes the number density of dark matter (in thermal equilibrium) and $H(T)$ is the Hubble expansion rate. %$\sigma$ stands for the cross section of DM pairly annihilating into all possible final states. $v$ is the relative velocity of two initial-state non-relativistic DMs in the CM frame and
$\langle\sigma v\rangle$ denotes the thermal averaged annihilation cross section of two DM. In the early universe,~$n = n^{eq}$, DMs were in the equilibrium. With expansion of the universe, the number density $n$ has been decreasing and DM froze out of the thermal equilibrium when the Hubble expansion rate became larger than its reaction rate.

More generally, when coannihilations become important, there are several particle species $i$ with
different masses and internal degree of freedom $g_i$. Each specie has its own number density $n_i$ and equilibrium number density $n^{ eq}_i$. In this case, the rate Eq.(\ref{Beq}) still applies, provided $n$ ($n^{ eq}$) is interpreted as the total (equilibrium) number density $n=\sum^N_{i=1}n_i$ ($n^{eq}=\sum^N_{i=1}n^{eq}_i$). Then, one can have the following evolution equation,
\begin{equation}\label{Beq2}
\frac{dn}{dt} = -3\emph{H}n - \sum^N_{i,j=1}\langle\sigma_{ij}\upsilon\rangle[n_in_j-n^{eq}_in^{eq}_j]
\end{equation}
Define
\begin{equation}\label{ri}
r_i \equiv n^{eq}_i/n^{eq} = \frac{g_i(1+\Delta_i)^{3/2}e^{-x\Delta_i}}{g_{eff}}
\end{equation}
where
\begin{equation}\label{delta}
\Delta_i=(m_i-m_1)/m_1, x=m_1/T
\end{equation}
and
\begin{equation}\label{geff}
g_{eff}=\Sigma^N_{i=1}g_i(1+\Delta_i)^{3/2}e^{-x\Delta_i}
\end{equation}
With the above definitions and the condition $n_i/n \approx n^{eq}_i/n^{eq}$, Eq.(\ref{Beq2}) can be written as
\begin{equation}\label{Beq3}
\frac{dn}{dt} = -3\emph{H}n - \langle\sigma_{eff}\upsilon\rangle[n^{2}-(n^{eq})^{2}]
\end{equation}
where the effective cross section is given by
\begin{equation}\label{sigma}
\langle\sigma_{eff}v\rangle=\sum^N_{i,j=1}\langle\sigma_{ij}v\rangle r_ir_j=\sum^N_{i,j=1}\langle\sigma_{ij}v\rangle \frac{g_ig_j}{g^2_{eff}}(1+\Delta_i)^{3/2}(1+\Delta_j)^{3/2}e^{-x(\Delta_i+\Delta_j)}.
\end{equation}
For the stau coannihilation, only the lightest bino-like neutralino and stau are involved in the calculation of Eq.(\ref{sigma}). When their mass difference is small enough, the annihilation cross sections of co-annihilating partners may contribute to $\langle\sigma_{eff}v\rangle$ significantly.
In particular, the coupling of stau with the SM Higgs boson can be greatly enhanced by the large mixing parameter, which is given by
\begin{equation}\label{interaction term}
{\cal L} \supset -\frac{m_\tau}{v}(A_\tau-\mu\tan \beta)(\tilde{\tau}^*_L\tilde{\tau}_Rh+\tilde{\tau}^*_R\tilde{\tau}_Lh).
\end{equation}
With this in mind, one can obtain the annihilation cross section of co-annihilating partners, for example the process $\tilde{\tau}_R \tilde{\tau}^*_R\rightarrow hh$,
\begin{eqnarray}
\langle\sigma v\rangle (\tilde{\tau}_R \tilde{\tau}_R^* \rightarrow h h )\simeq   \frac{1}{128\pi m^2_{\tilde{\chi}}} \left( \frac{m_\tau}{v} \right )^4   \frac{(A_\tau-\mu \tan \beta)^4}{m^4_{\tilde{\tau}_L}}.
\end{eqnarray}
This clearly shows such an annihilation process can become dominant when $|A_\tau-\mu\tan \beta|  \gg  m_{\tilde{\tau}_L}, m_{\tilde{\tau}_R} $. Similarly, the processes $\tilde{\tau}_R \tilde{\tau}^*_R\rightarrow WW/ZZ$ would also be enhanced due to the longitudinal contribution.

As mentioned above, a large mixing of staus may induce the vacuum instability. The classical stability of the electroweak-breaking vacuum requires that the smaller eigenvalue of the stau mass matrix is positive. However, as long as the left-right mixing is large enough, the scalar potential still has possibility to develop a global minimum different from the electroweak-breaking minimum, even though the classical condition is satisfied. We use the program \textsf{Vevacious}~\cite{Camargo-Molina:2013qva} to find the global minima of one-loop effective potential and determine whether the vacuum is stable or not. Firstly, \textsf{Vevacious} inputs the minimization conditions of the tree-level potential for the program \textsf{HOM4PS2}~\cite{Lee2008} to find all tree-level minima. Then these minima are used as starting points for gradient-based minimization (by Minuit~\cite{James1975} through PyMinuit~\cite{Pivarski}) of full one-loop potential with thermal corrections at a given temperature. If a minimum with lower potential energy than the desired symmetry-breaking (DSB) vacuum, \textsf{CosmoTransitions}~\cite{Wainwright:2011kj} is then called to calculate the tunneling time from the false DSB vacuum to the true vacuum. The generic expression for one-loop effective potential energy function used in \textsf{Vevacious} is given by
\beq\label{one-loop}
U^{1-loop}=U^{tree} + U^{counter} + U^{mass}
\eeq
where
\beq\label{Utree}
U^{tree}=\lambda_{ijkl}\phi_i\phi_j\phi_k\phi_l + A_{ijk}\phi_i\phi_j\phi_k + \mu^2_{ij}\phi_i\phi_j
+ {\rm constant~~terms}
\eeq
Here $\phi_i$ is the real scalar (a complex scalar field can be written as two separate real scalars) in four space-time dimensions. $U^{counter}$has a polynomial of the same degree in the same fields as $U^{tree}$ except that its coefficients are the renormalization dependent finite parts of the appropriate counterterms~\cite{Sher:1988mj,Degrassi:2012ry,Kobakhidze:2013tn,Kobakhidze:2014xda}. For a given field configuration $\Phi$, the $U^{mass}$ term has the form
\beq\label{vmass}
U^{mass}=\f{1}{64\pi^2}\sum\limits_n\{(-1)^{2s_n}(2s_n+1)(m^2_n(\Phi))^2[\ln(m^2_n(\Phi)/Q^2)-k_n]\}
\eeq
where $m^2_n(\Phi)$ are field-dependent squared masses and $s$ is the spin of involved field. The index \emph{n} runs over all of the real scalars, fermions, and vector degrees of freedom in the theory. \emph{Q} is the renormalization scale and $k_n$ is constant that depend on the details of the renormalization scheme. In the $\overline{DR}'$ scheme~\cite{Martin:2001vx,Jack:1994rk}, $k_n$ is 3/2 for all degrees of freedom while in the $\overline{MS}$ scheme, $k_n$ is 3/2 for scalars and Weyl fermions, but 5/6 for vectors.

The expression for the decay rate per unit volume $\Gamma/V$ for a false vacuum at zero temperature is given as~\cite{Coleman:1977py,Callan:1977pt}
\beq\label{decay rate}
\Gamma/V=Ae^{-S_4}
\eeq
where $\emph{A}$ is a factor which is related to the ratio of eigenfunctions of the determinants of the action's second functional derivative. $S_4$ is four dimensional bounce action that mainly contributes to the decay rate of the false vacuum. At zero temperature the $\emph{O}$(4) symmetric Euclidean action is written as
\beq\label{S4}
S_4=2{\pi}^2\int^\infty_0 d \rho \,{\rho}^3\[\f{1}{2}\sum^m_{i=1}\(\f{d\phi_i}{d\rho}\)^2+U(\phi_i(\rho))\]
\eeq
where $\rho$ is a radial coordinate in four-dimensional Euclidean spacetime and $m$ denotes the total number of scalar fields. $U(\phi_i(\rho))$ is the one-loop effective potential of scalar fields. The equation of motion and boundary conditions are
\beqa\label{S4eomandbc}
&&\f{d^2\phi_i}{d \rho^2} + \f{3}{\rho}\f{d\phi_i}{d\rho} = \f{\partial U(\phi_i)}{\partial \phi_i},\\
&&\lim_{\rho\rightarrow\infty} \phi_i(\rho) = \phi_i^f, \f{d\phi_i}{d\rho}\Big|_{\rho=0} = 0
\eeqa
where $\phi^f_i$ is the value of the scalar field $\phi_i$ at the false vacuum. Due to the suppression of $e^{-S_4}$, $\emph{A}$ is much less important than $S_4$ for the decay rate. If $\emph{A}$ is assumed to be about $(100{\rm GeV})^4$, $S_4$ must be at least 400 in order to make $\Gamma/V$ be roughly the age of the known universe to the fourth power. For a given path through the field configuration space from the false vacuum to the true vacuum, one can solve the equations of motion for a bubble of true vacuum that has critical size in an infinite volume of false vacuum~\cite{Coleman:1977py,Wainwright:2011kj}. Then one can calculate the bounce action and the tunneling time. When the temperature is sufficiently high, the dominant contribution to the decay rate comes from solitons that are $\emph{O}(3)$ cylindrical in Euclidean space rather than O(4) spherical. With the main thermal contributions, the equation of the decay rate per unit volume is changed into the following form
\beq\label{decaywidth2}
\Gamma(T)/V(T)= A(T)e^{-S_3(T)/T},
\eeq
where $\emph{T}$ is the universe temperature and $S_3$ is bounce action integrated over three dimensions which is defined as
\beq\label{S3}
S_3(T)=4{\pi}\int^\infty_0 d r \,r^2\[\f{1}{2}\sum^m_{i=1}\(\f{d\phi_i}{dr}\)^2+U(\phi_i(r),T)\]
\eeq
where $r$ is a radial coordinate in three-dimensional Euclidean space and $m$ represents the total number of scalar fields. $U(\phi_i(r),T)$ denotes the temperature dependent one-loop effective potential of scalar fields. The equation of motion yields
\beq\label{S3EOM}
\f{d^2\phi_i}{d r^2} + \f{2}{r}\f{d\phi_i}{dr} = \f{\partial U(\phi_i,T)}{\partial \phi_i}
\eeq
with the boundary conditions,
\beqa\label{S3bc}
\lim_{r\rightarrow\infty} \phi_i(r) &=& \phi_i^f,\\
\f{d\phi_i}{dr}\Big|_{r=0} &=& 0
\eeqa
where $\phi^f_i$ is the value of the field $\phi_i$ at the false vacuum. The non-tunneling probability $P(T_i,T_f)$ between the temperature $T_i$ and $T_f$ is given by
\beqa\label{probability}
P(T_i,T_f)= \exp\(-\int^{T_f}_{T_i} d T \,\f{dt}{dT}V(T)A(T)e^{-S_3(T)/T}\) .
\eeqa
 The surviving probability is calculated by looking for optical temperature $T_{opt}$, which appears at the minimum tunneling probability between the vacuum evaporation temperature $T_0$ and the starting temperature of color/charge vacuum breaking $T_{crit}$. More details of this method is described in~ \cite{Camargo-Molina:2014pwa}. In our numerical calculations, we implement \textsf{Vevacious-1.2.02} interfering with \textsf{CosmoTransitions-2.0a2}  to calculate the tunneling rate for our parameter space.

\section{Results and discussions}\label{section3}

In our study, the relevant model parameters come from soft SUSY-breaking parameters of stau sector, $\tan\beta$, bino mass parameter $M_1$ and higgsino mass parameter $\mu$. We fix $\tan\beta=50$ and take a common mass parameter $M_{SUSY}=5$ TeV for other irrelevant sparticle masses and pseudo-scalar mass $M_A$ for simplicity. In order to achieve the coannihilation efficiently, we adjust the bino mass parameter $M_1$ to require the dark matter not overabundant(usually the mass difference lies in the range of 10 GeV). We use spectrum generator SPheno-4.0.3~\cite{Porod:2003um} to produce SLHA files and Micromegas-5.0.4~\cite{Belanger:2001fz} to calculate DM relic density. There is no tachyon allowed in our spectrum. Also, we require our samples to satisfy the constraints of Higgs mass (122 GeV$ \leqslant m_h \leqslant$ 128 GeV) and DM relic density ($\Omega h^2 \leqslant 0.12$ ).
According to the stability condition, we can divide the parameter points into four categories. (1) Stable. There are no deeper charge-breaking minima developed; (2) Unstable. Points would tunnel out of the false DSB vacuum in three giga-years or less, (3) and rest as Long-lived; (4) Long-lived but thermally excluded. The DSB vacuum is long-lived but can become unstable if thermal corrections are included.

\begin{figure}[htp]
\centering
\includegraphics[width=2in,height=2.2in]{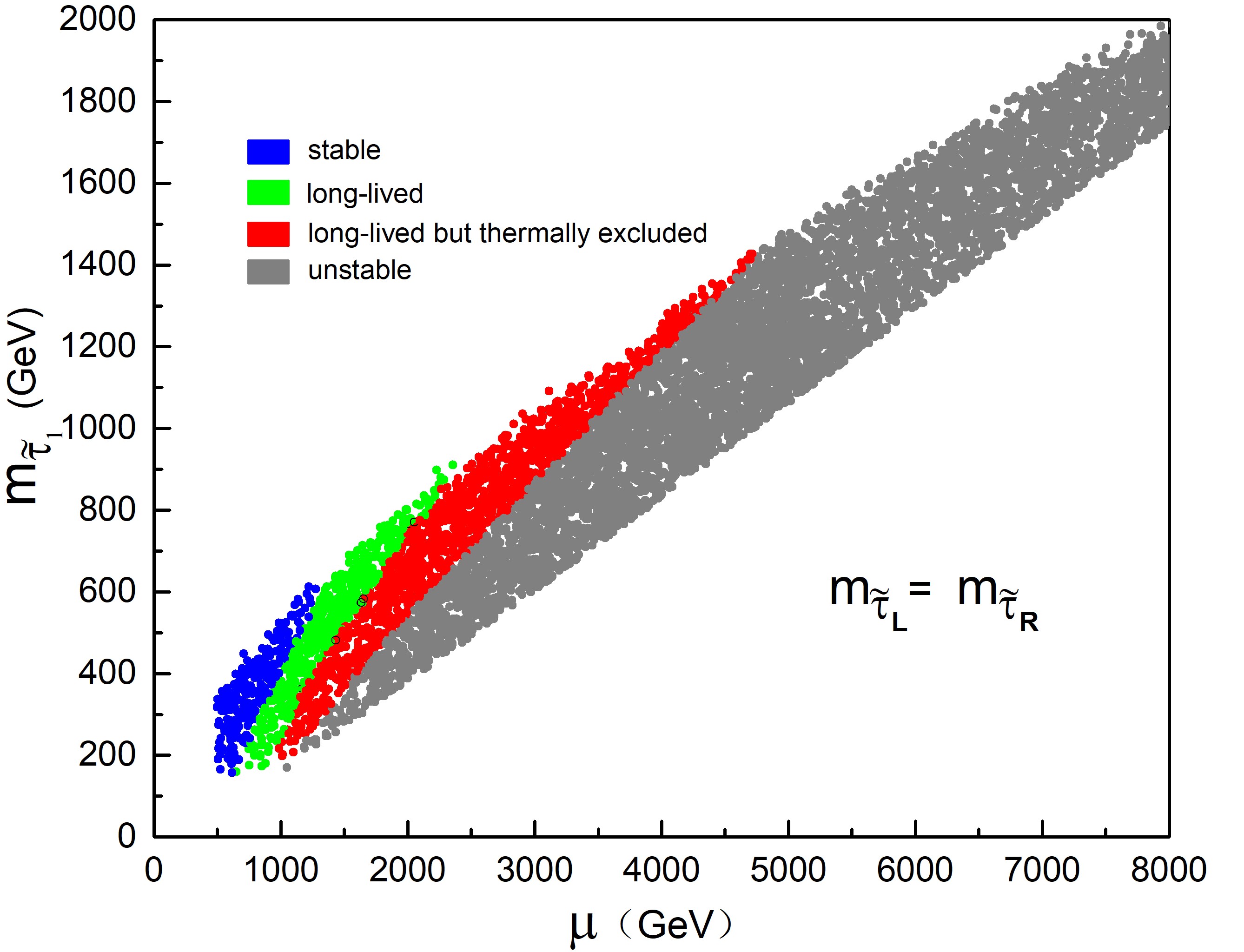}
\includegraphics[width=2in,height=2.2in]{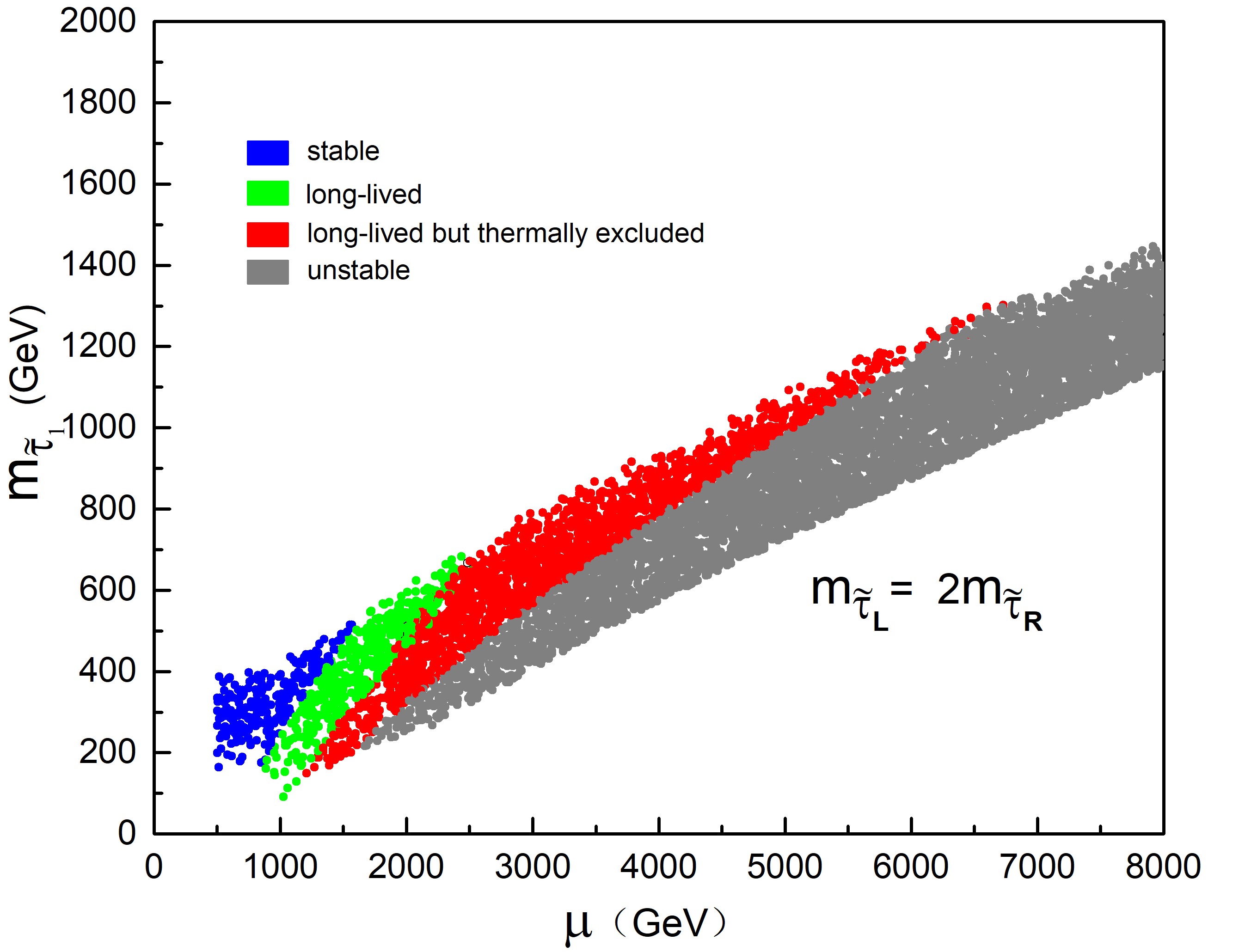}
\includegraphics[width=2in,height=2.2in]{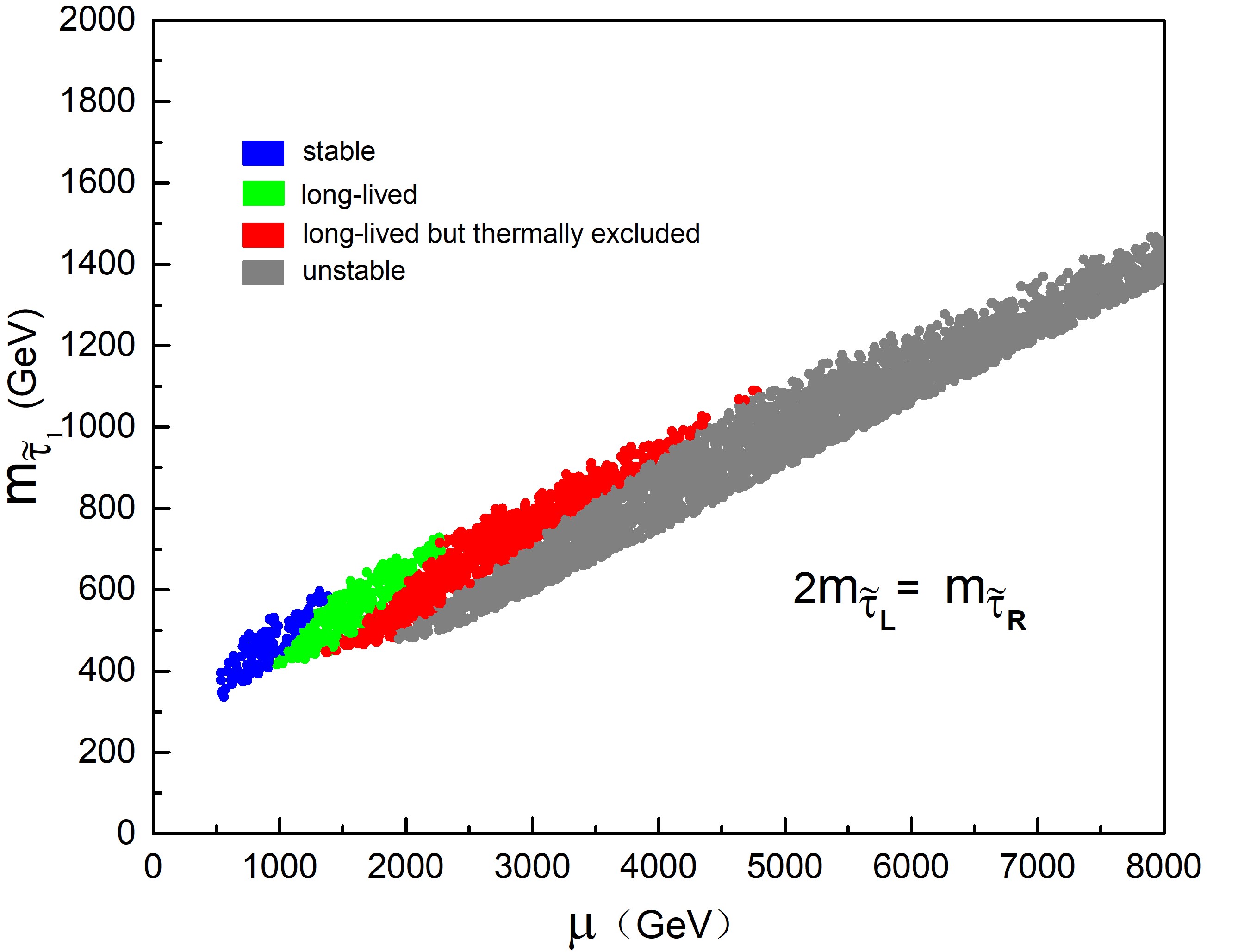}
\caption{The samples satisfying the constraints of Higgs mass and DM relic density are projected on the plane of $m_{\tilde{\tau}_1}$ and $\mu$ for $m_{\tilde{\tau}_{L}}=m_{\tilde{\tau}_{R}}$,~$m_{\tilde{\tau}_{L}}=2m_{\tilde{\tau}_{R}}$ and $2m_{\tilde{\tau}_{L}}=m_{\tilde{\tau}_{R}}$. The colormap denotes the different vacuum stability conditions.}
\label{stability}
\end{figure}
In Fig.~\ref{stability}, we show the samples satisfying the constraints of Higgs mass and DM relic density on the plane of $m_{\tilde{\tau}_1}$ and $\mu$. We compare the result of maximal mixing of staus  ($m_{\tilde{\tau}_{L}}=m_{\tilde{\tau}_{R}}$) with other two non-maximal mixing ($m_{\tilde{\tau}_{L}}=2m_{\tilde{\tau}_{R}}$, $2m_{\tilde{\tau}_{L}}=m_{\tilde{\tau}_{R}}$). We vary the mass parameters $m_{\tilde{\tau}_{L}}$ and $\mu$ within the ranges:  100 GeV $<m_{\tilde{\tau}_{L}}<$ 3 TeV and 100 GeV $<\mu<$ 10 TeV and $\tan \beta$ is fixed to be 50. It can be seen that the vacuum stability gives a strong bound on the parameter space of stau coannihilation.When the vacuum is required to be long-lived without thermal corrections, the lighter stau mass can be up to about 1.4 TeV for the maximum mixing $m_{\tilde{\tau}_{L}}=m_{\tilde{\tau}_{R}}$. If requiring the vacuum is long-lived and not thermally excluded, one can obtain the upper bound on the lighter stau mass $m_{\tilde{\tau}_1}\lesssim$ 900, 700, 750 GeV for $m_{\tilde{\tau}_{L}}=m_{\tilde{\tau}_{R}}$, $m_{\tilde{\tau}_{L}}=2m_{\tilde{\tau}_{R}}$ and $2m_{\tilde{\tau}_{L}}=m_{\tilde{\tau}_{R}}$, respectively. This is because the interaction between stau and LSP is enhanced most for maximal mixing in three cases so that the corresponding main annihilation channel $\tilde{\tau}_1 \tilde{\tau}^*_1 \to hh$ has larger cross section than other two cases and allows for heavier stau mass. Besides, it should be mentioned that the thermal correction plays an important role in the vacuum stability. The reason is  that for contribution from solitons that are O(3) cylindrical, it mainly comes from the states which have enough energy to easily pass the barrier, but it is usually suppressed by boltzmann factor $e^{-E/T}$. Only for a sufficient high temperature, this contribution would become important. We checked the best transition temperature and found it as high as few hundreds GeV comparing with the height of barrier $\sim$TeV$^4$. Finally we also checked our results at zero temperature with the fitting formula from \cite{Endo:2013lva} and found that they are well consistent.
%\begin{eqnarray}
%|m^2_{\tilde{l}_LR}|  \simeq  && \eta_l  [ 1.01\times 10^2 {\rm GeV} \sqrt{m_{\tilde{l}_L} m^2_{\tilde{l}_R}}-2.27 \times 10^4 {\rm GeV}^2
%+1.01 \times 10^2 {\rm GeV}(m_{\tilde{l}_L}+ 1.03 m_{\tilde{l}_R})   \nonumber\\
%&&+ \frac{2.97 \times 10^6 {\rm GeV}^3 }{m_{\tilde{l}_L}+m_{\tilde{l}_R}}
%-1.14 \times 10^8 {\rm GeV}^4(\frac{1}{m^2_{\tilde{l}_L}}+ \frac{0.983}{m^2_{\tilde{l}_R}})  ] ,
%\end{eqnarray}
%where $\eta_l$ varies from 0.9 to 1.0. We find our region of zero temperature long-lived region is well fit their formula with a $\eta_l\sim 0.9$.

\begin{figure}[htp]
\centering
\includegraphics[width=2in,height=2.2in]{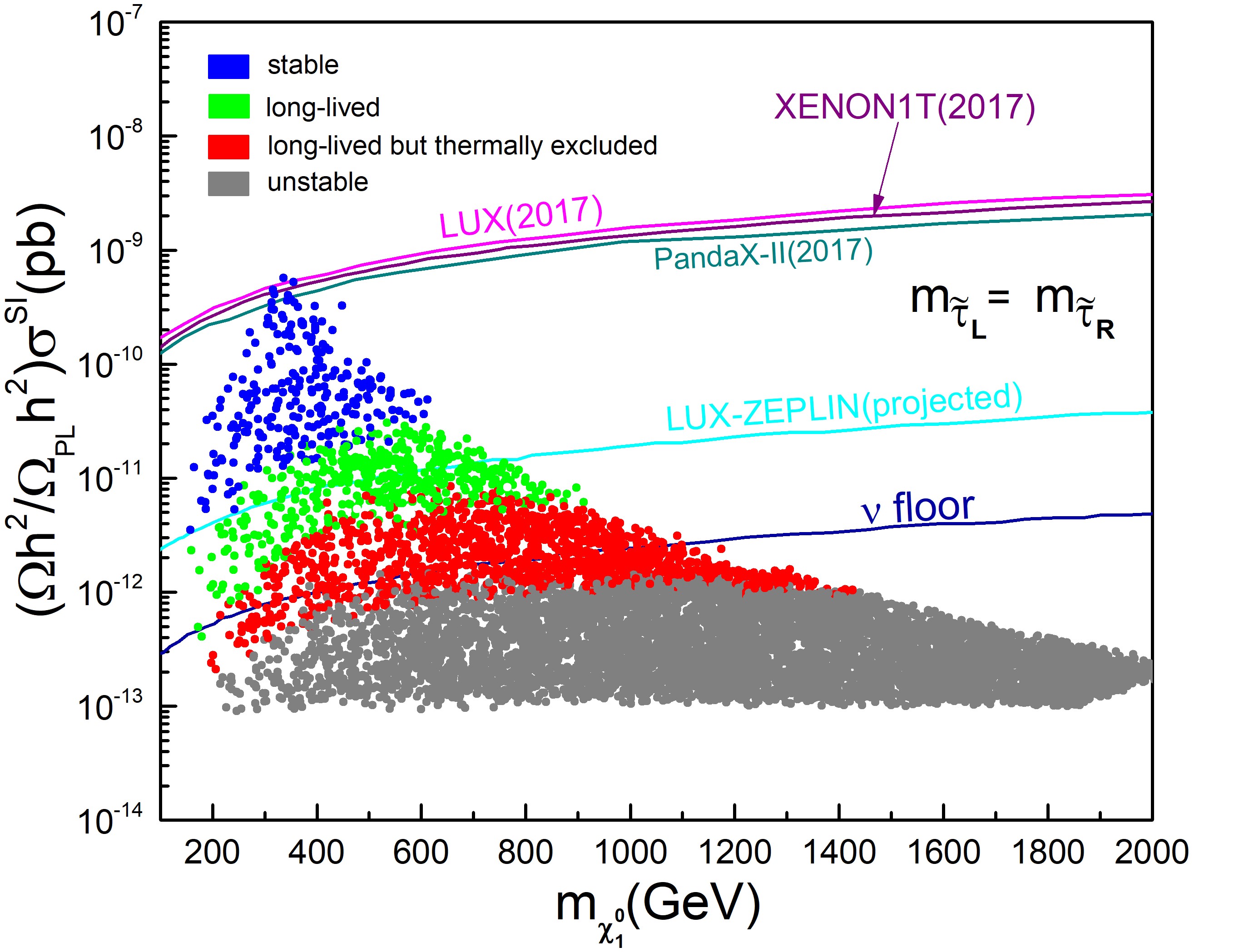}
\includegraphics[width=2in,height=2.2in]{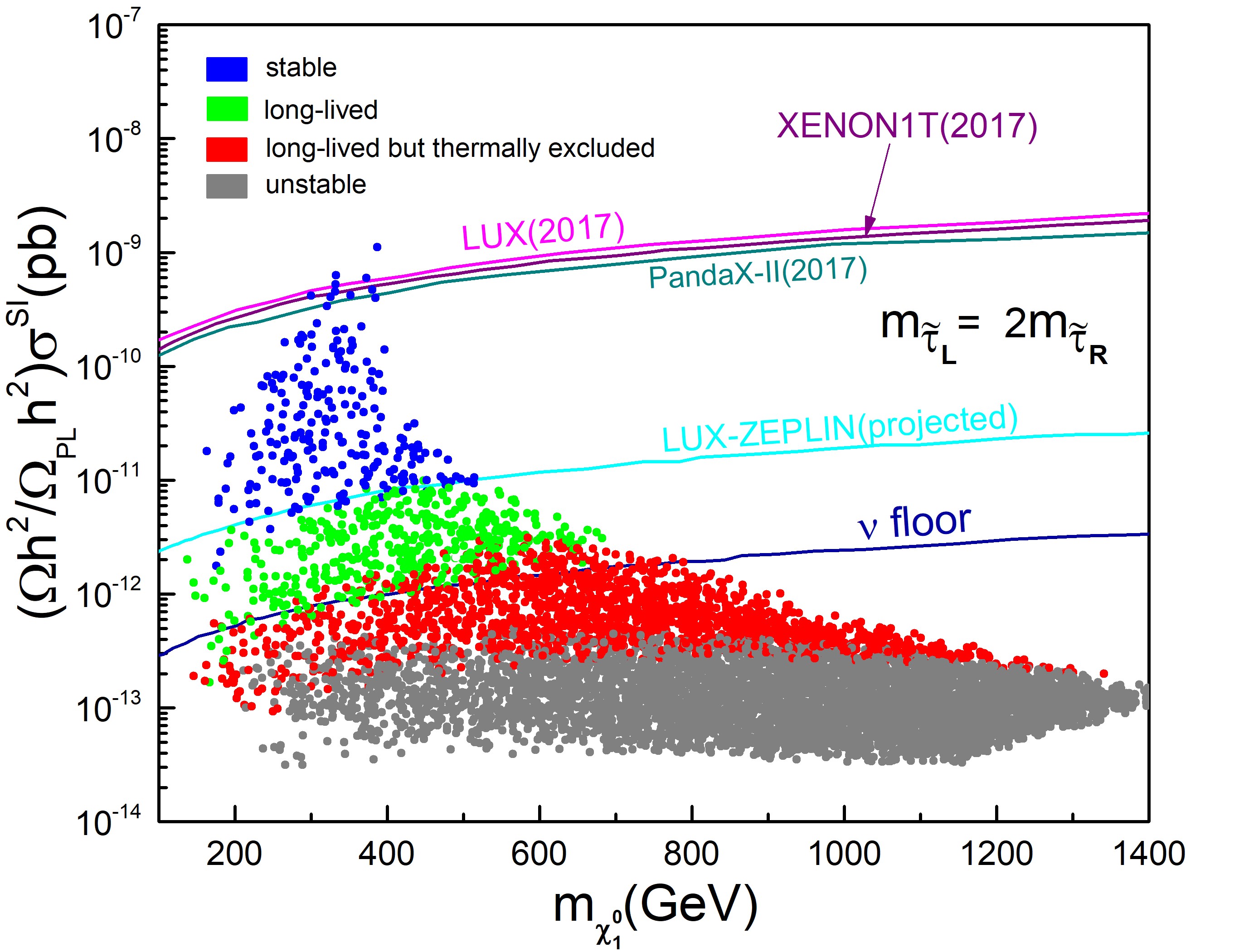}
\includegraphics[width=2in,height=2.2in]{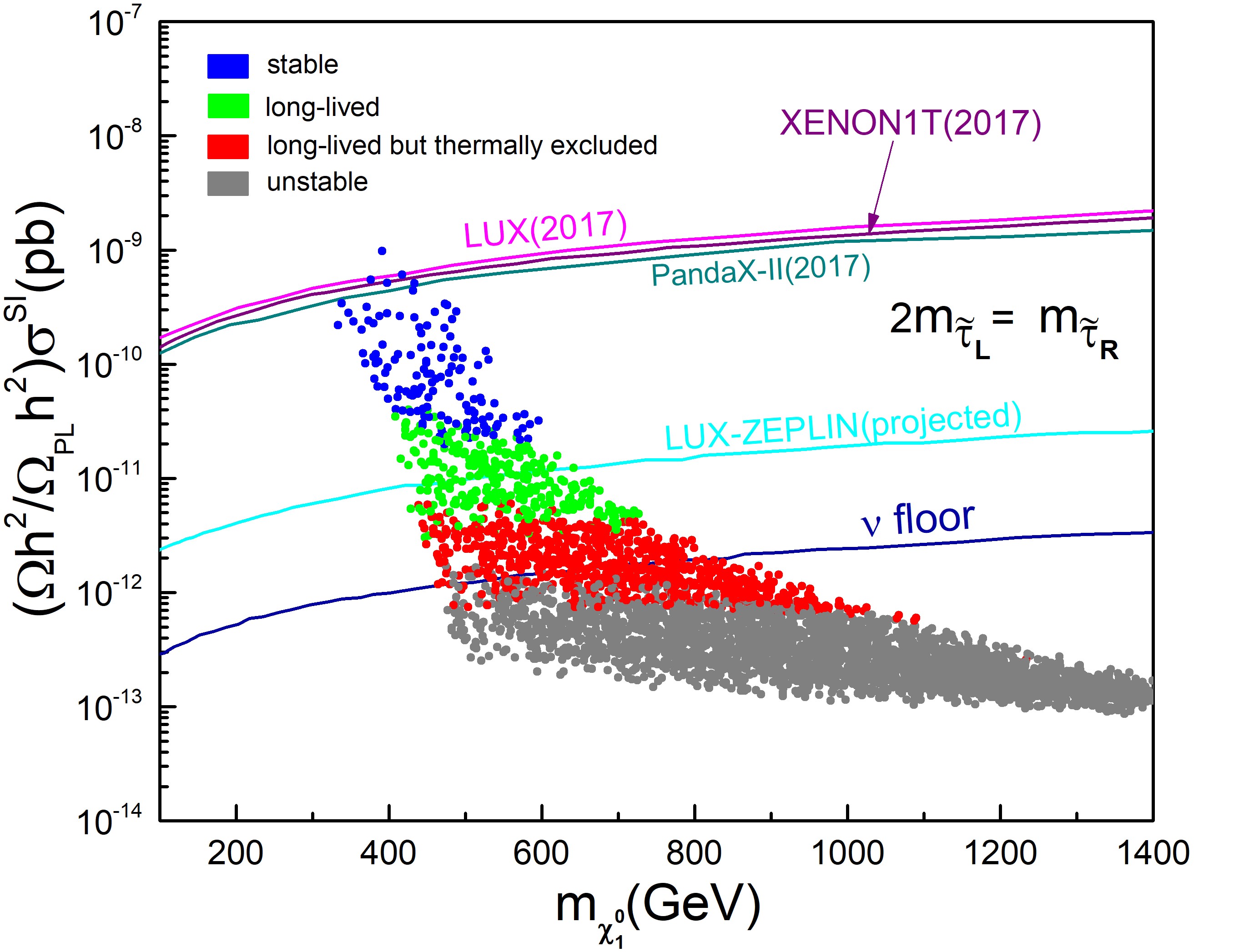}
\caption{Same as Fig.~\ref{stability}, but projected on the plane of $(\Omega h^2/\Omega_{PL} h^2)\sigma^{SI}$ and $m_{\chi^0_1}$. The observed 90\% C.L. upper limits from PandaX-II(2017) \cite{Yang:2016odq}, XENON1T(2017) \cite{Aprile:2017iyp}, LUX (2017) \cite{Akerib:2017kat} and the projected LUX-ZEPLIN's sensitivity are plotted~\cite{Akerib:2018lyp}.}
\label{dd}
\end{figure}
In Fig.~\ref{dd}, we calculate the spin-independent LSP-nucleon scattering cross section ($\sigma^{SI}$) for all above samples. We note that the thermal DM relic density can be inadequate in our stau coannihilation. For that case, the scattering cross section $\sigma^{SI}$ then must be rescaled by a factor of $\Omega_{\tilde{\chi}^0_1}h^2/\Omega_{PL}h^2$, where $\Omega_{PL}h^2$ is the relic density measured by Planck satellite \cite{Aghanim:2018eyx}.
%where the form factors of proton and neutron are taken as,
%\beqa\label{form factors}
%f^p_u\approx0.020,f^p_d\approx0.026,f^p_s\approx0.13 \nn \\ f^n_u\approx0.014,f^n_d\approx0.036,f^n_s\approx0.13
%\eeqa
From Fig.~\ref{dd}, we can see that almost all samples can escape the existing limits from PandaX-II(2017), XENON1T(2017) and LUX(2017). The projected LUX-ZEPLIN experiment will be able to improve the current sensitivity of LUX by about two orders of magnitude, and thus cover all of stable samples. On the other hand, it is worth to noticing that there are plenty of points below the neutrino floor, which is beyond the sensitivity of direct detections. Fortunately, all these points can be excluded by the constraint of vacuum stability. We note here  that although only the $\mu \tan \beta $ appears in the calculation of the dark matter relic and vacuum instability, the dark matter search results mainly relies on the value of Higgsino mass parameter $\mu$ and its mixing with bino. For a smaller $\tan \beta$, the maximum of the dark matter mass is essentially not  changed, however, the dark matter searches results can be much relaxed if we have a much larger $\mu$ parameter.

At last, one may wonder that the Higgs decaying into diphotons may be enhanced by the loop contribution from stau. We calculated the higgs effective coupling with photon $\kappa_\gamma$ ( $\kappa_\gamma=1$ for SM) and found most of surviving samples have a $\kappa_\gamma < 1.02$, which is much beyond near future LHC or lepton collider searches, so a few TeV lepton collider is needed to totally cover these parameter region.

%Before enclose section, we want to add following comments,
%\begin{enumerate}
%\item[(i)] Since only $\mu \tan\beta$ appears in the calculation of the dark matter annihilation process and vacuum stability, for a moderate $\tan\beta$, the maximum mass of dark matter is robust and independent of $\tan\beta$. However, the dark matter direct searches could be much weakened for a smaller $\tan \beta$ with a huge $\mu$.
%\item[(ii)] For the case of one of the stau decouples, the maximum dark matter mass is around 400 GeV. In this case, $\mu$ could also be very large than escape the dark matter direct searches.
%\end{enumerate}

\section{Conclusion}\label{section4}
In this work, we studied the vacuum stability constraint on the stau-neutralino coannihilation
in the MSSM. With the increase of the LSP mass, the observed relic density will produce an upper
limit on the mass of LSP and also its co-annihilating partner stau. We noticed that the main
annihilation channel of stau coannihilation is $\tilde{\tau}_1 \tilde{\tau}^*_1 \to hh$
in the parameter space with a large mixing of staus, which can relax the upper bound
on the stau mass obtained from the annihilation channel $\tilde{\tau}_1 \tilde{\tau}^*_1 \to f\bar{f}$.
Under the constraint of DM relic density, we found that the lighter stau mass should be less than
about 900 GeV to guarantee the vacuum stability. Besides, we noted that the vacuum stability
can play a complimentary role in probing the stau coannihilation scenario as to direct detections.

\section*{Acknowledgement}
LW thanks Archil Kobakhidze for helpful discussions. This work was supported by the National Natural
Science Foundation of China (NNSFC) under grant No. 11851303, 11705093 and 11675242,
by Peng-Huan-Wu Theoretical Physics Innovation Center (11747601),
by the CAS Center for Excellence in Particle Physics (CCEPP),
by the CAS Key Research Program of Frontier Sciences,
by a Key R\&D Program of Ministry of Science and Technology under number 2017YFA0402200-04,
and by the World Premier International Research Center Initiative (WPI Initiative), MEXT, Japan.

\end{document}